# Economic inequality and Islamic Charity: An exploratory agent-based modeling approach

Hossein Sabzian[1, *]   Alireza Aliahmadi[1]   Adel Azar[2]   Madjid Mirzaee[3]

1- Iran University of Science and Technolog
2- Tarbiat Modares University
3- Khajeh Nasir Toosi University of Technology

*Corresponding authors. E-mail: hossein_sabzian@pgre.iust.ac.ir

## Abstract

Economic inequality is one of the pivotal issues for most of economic and social policy makers across the world to insure the sustainable economic growth and justice. In the mainstream school of economics, namely neoclassical theories, economic issues are dealt with in a mechanistic manner. Such a mainstream framework is majorly focused on investigating a socio-economic system based on an axiomatic scheme where reductionism approach plays a vital role. The major limitations of such theories include unbounded rationality of economic agents, reducing the economic aggregates to a set of predictable factors and lack of attention to adaptability and the evolutionary nature of economic agents. In tackling deficiencies of conventional economic models, in the past two decades, some new approaches have been recruited. One of those novel approaches is the Complex adaptive systems (CAS) framework which has shown a very promising performance in action. In contrast to mainstream school, under this framework, the economic phenomena are studied in an organic manner where the economic agents are supposed to be both boundedly rational and adaptive. According to it, the economic aggregates emerge out of the ways agents of a system decide and interact. As a powerful way of modeling CASs, Agent-based models (ABMs) has found a growing application among academicians and practitioners. ABMs show that how simple behavioral rules of agents and local interactions among them at micro-scale can generate surprisingly complex patterns at macro-scale. In this paper, ABMs have been used to show (1) how an economic inequality emerges in a system and to explain (2) how sadaqah as an Islamic charity rule can majorly help alleviating the inequality and how resource allocation strategies taken by charity entities can accelerate this alleviation.

# 1. Introduction

Many of social phenomena like residential segregation, norm formation, technology diffusion, and evolution of states occur not due to separate choices by constituent individuals but mainly because of a networked interaction among constituent individuals over time. As a matter of fact, such phenomena have a nature entirely different from their constituents. Modeling the formation of these collective phenomena has been a great target for mainstream socio-economic modeling approach but it has not captured it sufficiently. This mainstream modeling approach which often called equation-based modeling (EBM) has been frequently used in different disciplines of social sciences. However, EBMs lack a needed functionality in explaining how the interactions among micro-components of a system can cause an interestingly different macro-behavior for that system. In fact, they perform very poorly in modeling the emergent properties of real-life systems, namely how a whole arises from the interactions among its simpler and lower-level parts so that it exhibits properties that its simpler and lower-level parts can never exhibit. For tacking such a limitation, agent based modeling (ABM) has been developed. ABM is a kind of computational model which explores systems of multiple interacting agents which are spatially situated and evolve over the time. ABMs are highly effective in explaining how complex patterns emerge from micro-level rule during a period of time. In contrast to EBMs that are based on deductive reasoning, ABMs properly work not only as an inductive reasoning technique where a conclusion is formed from a series of observations but also as a pure form of abductive reasoning where the best explanation for the phenomena under study is inferred via simulation. ABMs have become a major modeling trend in a large number of domains ranging from spread of epidemics (Situngkir, 2004) and the threat of bio-warfare (Caplat, Anand, & Bauch, 2008) to formation of norms (Axelrod, 1986), supply chain optimization (Van Dyke Parunak, Savit, & Riolo, 1998) and collaboration in project teams (Son & Rojas, 2010). In contrast to EBMs which majorly focus on relationship among macro-variables of a system in top-down manner, ABMs try to model how local and predictable interactions among micro-components of a system can generate a complex system-level behavior (Macy & Willer, 2002). ABM methodology is rooted in complexity theory and network science. In terms of complexity theory, ABMs are developed to explain how simple rules generate complex emergence (i.e. a process model) and in terms of network science ABMs are used to analyze the pattern that arise from agents' interactions over the time (i.e. a pattern model) (Wilensky & Rand, 2015).

In this paper, we want to explore a novel way to describe and analyze the economic inequality by simulating the behavior of agents in an ABM. Also, to answer this problem (economic inequality), we propose the concept of sadaqah as a charity tool which is frequently emphasized in Islam and can play a vital role in reducing economic inequality. Additionally, the way the resource allocation strategies of charity entities can influence the economic inequality are also discussed. The rest of this paper is organized as follows: the section 2 deals with the economy as a complex adaptive system (CAS). The issue of inequality is discussed in section 3 and its drivers and measurement methods are presented. Section 4 is concerned with inequality and charity and the Islamic viewpoint of charity is discussed in section 5. CAS and ABMs are elucidated in section 6. Section 7 deals with specifications of systems under study and discussion of simulation results. The conclusion is provided in section 8.

## 2. The Economy as a Complex Adaptive System

In the last two decades, a new way of thinking about systems has been emerged, essentially in the natural sciences. However, all the common physical laws are obeyed, at now, many systems are viewed as 'complex' and, per se, using prevalent approaches to theorizing and modeling cannot be helpful. In general, such systems so called self-organized structures that assimilate and disperse energy and, despite their visible complexity, in many cases can obey some quite ordinary behavioral rules in time and space. Although, this behavior cannot be expressed by theories that presume the actuality of historical time and, consequently, disregard the real processes that manifest within and beyond complex systems. For instance, it can be debated that the results of these historical processes cannot be regarded as the answers of constrained dynamic optimization problems. This brings up some essential questions for economists, given that all economic systems can be arranged as complex systems. In this regard, systems that absorb information from surrounding environment and accumulate knowledge that can help action are usually called 'complex adaptive systems'.

Now the question is that, is an economic system a complex adaptive one? The answer is positive if we consider four general properties for economic system as follows:

- Such a system is the build up from connections that are established between elements of systems that allow higher levels of accumulation at the emergence of organized complexity.
- It is a disperse structure that change energy into work and transforms information into knowledge for the aim of building, preserving and developing the organized complexity of the system.
- Such a system should be display some sort of structural invariability due to the intrinsic hierarchical and 'bonding' feature of the relations between constituents that are organized as structural development advances.
- The evolutionary process that such a system encounters can only be perceived in an explicit historical time dimension.

In recent years, the literature about complexity economics has been developed in so many areas. These include evolutionary models inspired by Nelson and Winter (Nelson & Winter, 1982) and Hodgson (Hodgson, 1998),William Brock and Stephen Durlauf's study of social interaction(Brock & Durlauf, 2001), study of firm size by Robert Axtell (Axtell, 2001), Alan Kirman and his colleagues models of financial markets (A. Kirman, Foellmer, & Horst, 2005) and the agent-based simulation of general equilibrium(Gintis, 2006a, 2006b).[1]

When we look carefully at actual world, constrained optimization models cannot capture the behavior of complex adaptive systems. This is an essential departure from the presumptions

---

[1] -For a comprehensive overview of computational methods in complexity economics, look at (Amman, Tesfatsion, Kendrick, Judd, & Rust, 1996; Tesfatsion, 2002)

existing in conventional economic theories. Such systems should be analyzed 'in' time and this limits the way that mathematics can be used. Standard economic theory includes the applying of an ahistorical body of logical clauses to display attitudes perceived in the historical domain. In opposition, complex adaptive system theory copes directly with the fundamental principles that rule the behavior of systems in history. Therefore, we can say that thinking about the economy and its sub-components as complex adaptive systems can allow us to abscond from of such scientific impasses.

In economic thought, Schumpeter's contributions toward the process of "creative destruction" conform to complex adaptive systems theory (Foster, 2001). In spite of the fact that these ideas have been displayed in different appearances by notable economists such as Herbert Simon, Friedrich von Hayek and Nicholas Georgescu-Roegen, Kenneth Arrow and Brian Arthur (which the last two ones centrally involved in the founding of the Santa Fe Institute in the late 1980s). In the conventional economic theories, there has been only limited interest in applying the complex adaptive systems theory.

# 3. Inequality in economics

One of the issues that concern most of economic and social policy makers across the globe is the worsening situation of economic equality[2] in recent years. In this regard, much of the debate has come down to two views: the first one is mainly concerned with the inequality of consequences in the material dimensions of welfare and that may be the result of conditions beyond one's control (ethnicity, family background, gender, and so on) and also effort and talent. The second viewpoint is concerned with the inequality of opportunities that it focuses only in the circumstances beyond one's control, that affect one's potential yields. Usually, inequality of outcomes takes place when individuals do not own the same level of wealth or overall living economic conditions.

## 3.1. Drivers of Inequality

By Looking at the studies and researches that carried out in recent years, it can be said that several factors have the pivotal role in formation of inequality that the most important ones mentioned as follows:

### 3.1.1. Technological change

Over the last two decades, we can observe notable leaps and bounds in productivity which played a key role in economic growth. This phenomenon originated from information technology improvement which also played a pivotal role in driving up the skill gap among labors, resulting in increased labor income inequality. This is because technological changes can raise the demand for skilled labor and capital over low-skilled labor by obliterating many jobs by means of automation or increasing the level of skills required to get or keep those jobs (Acemoglu, 1998; Card & DiNardo, 2002).

---

2 - Economic inequality sometimes refers to income inequality, wealth inequality, or the wealth gap.

### 3.1.2. Financial globalization

Globalization in financial area can assist better allocation of capital in international dimension and encourage international risk sharing. Meanwhile, increasing the flow of financial resources, especially foreign direct investment (FDI) have been exhibited to elevate inequality in income in both emerging market and advanced economies (Freeman, 2010). One potential description is the centralization of assets and liabilities in relatively higher technology-intensive sectors, which make upward pressure for wages of higher skilled workers.

In addition, Deregulation in financial sector and globalization have been proposed the factors that describe the increase in accumulation of wealth, wages and relative skill intensity in the finance industry, especially in advanced economies(Furceri & Loungani, 2013; Philippon & Reshef, 2012).

### 3.1.3. Financial deepening

Increasing the level of financial deepening potentially can decrease income inequality, while amending the allocation of resources(Dabla-Norris, Ji, Townsend, & Unsal, 2015). However, in theory expressed that the financial development could make rewards for the rich in the early stages of development, but in overall, as economies develop over the time, interests become more broadly shared (Greenwood & Jovanovic, 1990).

### 3.1.4. Changes in labor market institutions

Economic dynamism can be foster by higher level of flexibility in labor market institutions through reallocating resources to more productive firms. Although, greater flexibility can impose challenges for workers, mostly those with low skills, and therefore play a key role in explaining developments in inequality (Alvaredo, Atkinson, Piketty, & Saez, 2013). In addition, declining the number of trade union members in the last decades reduced the relative bargaining power of labor, deteriorated the wage inequality (Frederiksen & Poulsen, 2010; Wilkinson & Pickett, 2010).

### 3.1.5. Redistributive policies

Historically, inequality in advanced economies have relieved through doing actions by governments —such as establishment of social safety nets and progressive taxes(Office, 2011). Nevertheless, over the past few decades, the effective tax rates have declined in some advanced economies, which exacerbated the income gap significantly(Hungerford, 2013).

Some actions like tax and transfer systems play an essential role in decreasing income inequality. Transferring the cash – through pensions and unemployment benefits – describes more than seventy percent of overall redistributive impact, and taxes for the remaining part. Nevertheless, there are different situation among the OECD countries in the size, composition and progressivity about taxes and cash transfers(Joumard, Pisu, & Bloch, 2012). In addition, the taxation of capital income and wealth has been alleviated in many countries, which has visibly reduced the redistributive effect of tax systems. Actually, capital income tends to be centralized in the upper income or wealth deciles (Alvaredo et al., 2013; Frederiksen & Poulsen, 2010).

### 3.1.6. Education

Usually, so called that the Education sector playing a pivotal role in reducing inequality in income or wealth, as it explains career choice, achievement to jobs and the level of pay. Moreover, due to the existence of asymmetric information in job market, it plays a key role as a signal of capability and productivity of labors.

## 3.2. Methods for Measuring Inequality

Inequality is a broader concept than poverty in that it is defined over the entire population, and not just for the population below a certain poverty line. Some of the commonly used measures of inequality are discussed as following:

### 3.2.1. Gini coefficient of inequality

The most commonly used measure of inequality is the Gini coefficient. It is based on the Lorenz curve, an accumulative curve that checks the distribution form of a specific variable (e.g. income or wealth) with the uniform distribution that display equality.

For instance, if $x_i$ be a point on the X-axis, and $y_i$ a point on the Y-axis on the Lorenz curve, Then

$$Gini = 1 - \sum_{i=1}^{N} (x_i - x_{i-1})(y_i + y_{i-1})$$
(1)

### 3.2.2. Generalized Entropy measures

The GE index, such as the Atkinson index, is marked as a group of income or wealth inequality measures. The common formula is as follows:

$$GE(\beta) = \frac{1}{\beta(\beta-1)} \left[ \frac{1}{N} \sum_{i=1}^{N} \left( \frac{y_i}{\bar{y}} \right)^{\beta} - 1 \right]$$
(2)

Where in this formula, $\bar{y}$ is the average income (or wealth per capita). The quantities of GE measures fluctuate between zero and $\infty$, with zero displaying an equal distribution and in contrast, the higher value of GE represents the worsening of inequality.

The parameter $\beta$ in the GE relation displays the weight given to gaps between incomes at diverse parts of the income distribution.

### 3.2.3. Atkinson's inequality measures

Atkinson has proposed another class of inequality measures that are used from time to time. This class also has a weighting parameter e (which measures aversion to inequality) and some of its theoretical properties are similar to those of the extended Gini index.

$$A_e = 1 - \left[ \frac{1}{N} \sum_{i=1}^{N} \left( \frac{y_i}{\bar{y}} \right)^{1-e} \right]^{1/(1-e)} , e \neq 1$$

$$= 1 - \frac{\prod_{i=1}^{N} (y_i^{(1/N)})}{\bar{y}} , e = 1$$

(3)

### 3.2.4. Decile dispersion ratio

It is the ratio of the mean income of the wealthiest x percent of the population to the mean income or wealth of the poorest x percent. It declares the income (or income share) of the poor as a proportionate of the rich. Common decile ratios contain ratio of the income or wealth of the 10 per cent wealthiest to that of the 10 per cent poorest. However, the decile ratio is easily interpretable because it declines data about incomes in the middle of the distribution, and does not even apply information about the distribution of income inside of the top and bottom deciles.

## 4. Inequality and Charity

The literature about inequality represents an ambiguous relationship between inequality and charitable actions. On the one hand, the sociological literature spotlights on a negative relationship between inequality and unanimity. Because they believe that inequality, worsen social distance and causing to social separation. The lower interaction among people potentially declines the tendency to aid others because they think they have different attitude and it is natural (Durkheim, 1893; Paskov & Dewilde, 2012; Wilkinson & Pickett, 2010). In addition, usually there is a preference for income homophily in society, so that higher level of inequality reduces the eagerness to connect to social activities (Alesina & La Ferrara, 2000). On the other hand, pure philanthropy indicates that inequality brings up solidarity (Bowles & Gintis, 2000; Fehr & Schmidt, 1999) and, so, participation in charitable actions, particularly if inequality is the outcome of reduced incomes or wealth at the lowest part of the distribution (Charness & Rabin, 2002).

One instance is an amplified unemployment in recession period (Galbraith, 1998), which greatly hits low-skilled or unskilled laborers because of skill-biased technological change or wage competition from abroad (Bound & Johnson, 1992). In addition, impure philanthropy also signals about a positive connection between inequality and charity actions, for example by using the warm-glow theory of giving (Andreoni, 1988).

Based on a recent study that has been done by Mastromatteo and Russo (2017) there is a visible positive relevancy between inequality and charity: probability of actively involving in charitable actions is higher in countries with more inequality in income or wealth. Since charitable funds or organizations are usually considered as a successor to public services or perhaps a response to an inadequate level of them, therefore, there is a negative relationship between charity participation and existence of the generous welfare state (Mastromatteo & Russo, 2017).

Moreover, the dominant records of giving in history and economic theory represents that philanthropy is an offsetting force against economic inequality. In addition, the findings in this area show that the more unequal the income distribution, the larger the contribution share of rich people. If this explanation is accurate, then philanthropy has relieved income inequality, turning the increased wealth of the prosperous few into improvements in knowledge and facilities that increment the quality of life for all. Abigail Payne and Justin Smith (2015) discover that increases in level of inequality in Canadian neighborhood from 1991 to 2006 are positively correlated with total charitable giving, however with significant nonlinearities in income that affirm a different reaction for high earners. Nonetheless, the economic history of altruism and philanthropic measures is insignificant(Payne & Smith, 2015).

## 5. Islamic Viewpoint to Charity

The notable characteristic of the Islamic doctrine is the dynamic nature of redistribution. Indeed, Islam presents some sort of detailed rules and orders that explaining the boundaries imposed on the accumulation of wealth. The existence of some laws in Islam about inheritance and anti-usury are two characteristic examples.

Concerning the inheritance laws, the Holy Quran describes that two-thirds of one's wealth be assigned to different family members, including very distant relatives that making it a nearly classless distribution system (Kuran, 2008).

In Islam, doing of charitable actions is divided into voluntary (sadaqah) and obligatory (zakat) ones. The Holy Quran wants a believer with adequate economic means to donate a fraction of her piled up income for alms. Zakat is devoted among the poor, debtors, travelers, the zakat collector and for captives or slaves. The other way of giving in Islam is Sadaqah that mentions to the processes by which one authenticates something or reinforces it. Technically, offerings and alms in Islam are meant to work for as reinforcement and supplement to the success of prayer as a medication with a try to solve one problem or the other. In this study, we consider the sadaqah as a way for the funding of poor or needy agents and a solution to improve the inequality condition in economics. Actually, the purpose of this study is to simulate (1) how sadaqah can help reduce the economic inequality of a society and (2) by which resource allocation strategies, charity organizations can reduce the economic inequality very efficiently.

## 6. Agent based modeling and complex adaptive systems

Developed from the fields of complexity, cybernetics, cellular automata and computers, the Agent-Based Modeling (ABM) has gained lots of popularity in the 1990s and shows a growing migration not only from EBMs such as econometric models but also from the more classical simulation approaches such as the discrete-event simulation (Heath & Hill, 2010). ABMs have a wide range of application domains ranging from biological systems (Caplat et al., 2008; Situngkir, 2004) to engineered ones (Olfati-Saber & Murray, 2004). The primary reason for modelers who apply ABMs is its very high strength in modeling complex adaptive systems (CAS) in comparison with other modeling methods. As a specific type of system, a CAS includes some basic characteristics as following:

1- It is composed of a number of components usually called" agent"(Holland, 2002; Wilensky & Rand, 2015). These agents can be very heterogeneous.

2- Its agents interact with each other in a non-linear (non-additive) way (Holland, 2002; Wilensky & Rand, 2015).

3- Its agents can adapt or learn (Holland, 2002) so agents can experience and accumulate knowledge.

4- It is non-ergodic (Kauffman, 2000; Moss, 2008).Therefore, it is highly sensitive to initial conditions.

5- It self-organizes and its control is intensely distributed among its agents(Chan, 2001; Wilensky & Rand, 2015).

6- It exhibits emergence (Chan, 2001; Holland, 2002; Wilensky & Rand, 2015). It means, from the interactions of individual agents arises a global pattern or an aggregate behavior which is characteristically novel and irreducible to behavior(s) of agent(s).

7- Its agents can co-evolve and change the system's behavior gradually (Kauffman, 1992).

8- It shows "far from equilibrium" phenomenon(Nicolis & Prigogine, 1989). isolated systems have a high tendency towards equilibrium and this will cause them to die. The "far from equilibrium" phenomenon shows how systems that are forced to explore possibilities space will create different structures and novel patterns of relationships (Chan, 2001)[3].

9- It is time asymmetric and irreversible. One characteristic of a CAS is time asymmetry. Asymmetry in time occurs when a system passes a bifurcation point, a pivotal or decisional point where an option is taken over another or others, leading to time irreversibility. Irreversibility means that the system cannot be run backwards— rewound or reversed—so as to reach its exact initial conditions. Systems which, when run in reverse, do not necessarily or typically return to their original state are said to be asymmetric in time(Prigogine & Stengers, 1997), and asymmetry in time is important in testing for a complex adaptive system. If system-time is symmetric in both directions, then it is reversible, and it is not a CAS but a deterministic system. Complex adaptive systems are asymmetric in time, irreversible and nondeterministic. So, in a CAS one can neither predict nor "retrodict," even with infinite information on initial conditions, because the system "chooses" its forward path. Its "choice" is indeterminate, a function of statistical probability rather than certainty(Rogers, Medina, Rivera, & Wiley, 2005).

10- Distributed control: the behavior of a CAS is not controlled by a centralized mechanism, rather, it is completely distributed among its constituent parts. The interactions of these constituent parts cause a CAS to exhibit a coherent macro-level behavior(Chan, 2001).

The philosophy of agent-based modeling comes directly from the idea that a CAS can be effectively modeled and explained by creating agents and environment, characterizing behavioral rules of agents, and specifying interactions among them(Wilensky & Rand, 2015). Modeling a CAS needs a specific type of methodology. EBMs such as statistical modeling techniques lack a needed functionality for this purpose because they just decompose a system into its main parts and model the relationship among them (a top-down approach) while neglecting the fact that the system itself is entity beyond its constituent parts and it needs to be analyzed as an emergence of its constituent parts (a bottom-up approach).

---

[3] In thermodynamics, systems that does not have any exchange of energy and matter with their surrounding environment are called "isolated systems". Such systems have a tendency to evolve towards equilibrium. But, our surrounding is enriched by phenomena arising from conations far from equilibrium. Some examples can be turbulences, fractals and even life itself. (Jaeger & Liu, 2010).

## 6.1. Unique Characteristics of ABMs

EBM and ABM have stemmed from two distinct epistemological frameworks. The former is grounded on neoclassical economic theories (NET) where the issues such as unbounded rationality, perfect information, deductive reasoning and low-rate heterogeneity are discussed while the latter is built upon complexity theory where the issues such as bounded rationality, information asymmetry, network interaction, emergence and inductive reasoning are taken into consideration (CT)(Al-suwailem, 2008; Moss, 2008). This has made ABMs specifically privileged in modeling CASs. These privileges can be summed as following:

1. *Bounded rationality*. The environment in which agents interact is highly complex and unbounded rationality is not a viable assumption for it (Al-suwailem, 2008; Wilensky & Rand, 2015), agents have limited possibilities not only for receiving information but also for its processing. AB modelers contend that socio-economic systems have an inherently non-stationary nature, due to continuous novelty (e.g., new patterns of aggregated behavior) endogenously introduced by the agents themselves (Windrum, Fagiolo, & Moneta, 2007). Therefore, it is extremely difficult for agents to learn and adapt in such a turbulent and endogenously changing environments. On this basis, AB researchers argue that assumption of unbounded rationality is an unsuitable for modelling real world systems and agents should not only have bounded rationality but also adapt their expectations in different periods of time.
2. *Exhibition of emergence:* Since ABMs can model how micro-dynamics result in a high-level macro-dynamic they can be used as the best method for exhibiting emergent properties. On this basis, ABM does not require knowledge of the aggregate phenomena, in fact, researchers do not need to know what global pattern results from the individual behavior. When modeling an outcome variable with EBM, you need to have a good understanding of the aggregate behavior and then test out your hypothesis against the aggregate output (Wilensky & Rand, 2015)
3. *Bottom-up perspective*. A macro-system is an outcome of the way its sub-systems interact so the properties of macro-dynamics can only be properly understood as the outcome of micro-dynamics involving basic entities/ agents (Tesfatsion, 2002). This contrasts with the top-down nature of traditional neoclassical models, where the bottom level typically comprises a representative individual and is constrained by strong consistency requirements associated with equilibrium and unbounded rationality(Eliot R. Smith, 2007; Macy & Willer, 2002). Conversely, AB models describe strongly heterogeneous agents living in complex systems that evolve through time(A. Kirman, 2010; A. P. Kirman, 1997). Therefore, aggregate properties are interpreted as emerging out of repeated interactions among simple entities rather than from the consistency requirements of rationality and equilibrium imposed by the modeler (Dosi & Orsenigo, 1994)

4. *Heterogeneity and discrete nature*: an ABM can nicely model a heterogeneous population, whereas equational models typically must make assumptions of homogeneity. In many models, most notably in social science models, heterogeneity plays a key role. Furthermore, when you model individuals, the interactions and results are typically discrete and not continuous. Continuous models do not always map well onto real-world situations(Wilensky & Rand, 2015)

5. *Networked interactions*: Interactions among economic agents in AB models are direct and inherently non-linear (Fagiolo, 1998; Silverberg, Dosi, & Orsenigo, 1988). Agents interact directly because current decisions directly depend, through adaptive expectations, on the past choices made by other agents in the population (i.e. a widespread presence of externalities). These may contain structures, such as subgroups of agents or local networks. In such structures, members of the population are in some sense closer to certain individuals in the socio-economic space than others. These interaction structures may themselves endogenously change over time, since agents can strategically decide with whom to interact according to the expected payoffs. When combined with heterogeneity and bounded rationality, it is likely that aggregation processes are non-trivial and, sometimes, generate the emergence of structurally new objects (Lane, 1993a, 1993b).

6. *Comprehensiveness:* results generated by ABMs are more detailed than those generated by EBMs. ABMs can provide both individual and aggregate level detail at the same time. Since ABMs operate by modeling each individual and their decisions, it is possible to examine the history and life of any one individual in the model, or aggregate individuals and observe the overall results. This "bottom-up" approach of ABMs is often in contrast with the "top-down" approach of many EBMs, which tell you only how the aggregate system is behaving and do not tell you anything about individuals. Many EBMs assume that one aspect of the model directly influences, or causes, another aspect of the model, while ABMs allow indirect causation via emergence to have a larger effect on the model outcomes(Wilensky & Rand, 2015).

7. *Randomness and indeterminacy*: One important feature of agent-based modeling, and of computational modeling in general, is that it is easy to incorporate randomness into your models. Many equation-based models and other modeling forms require that each decision in the model be made deterministically. In agent-based models this is not the case; instead, the decisions can be made based on a probability(Wilensky & Rand, 2015).

## 6.2. **Main uses of ABMs**

ABMs can be used in *description and explanation*. Like all models, an AMB is a simplification of a real world system which doesn't entail all of its aspects so it is distinguishable from real world system and can help its understanding. The exploratory nature of ABM indicates that they can be used to pinpoint the essential mechanisms underlying the phenomena under study. a subject matter expert can use an AMB as a proof that his or her hypothesized mechanisms sufficiently account for the aggregate behavior under study. (Wilensky & Rand, 2015).

Explanation is strongly believed to be a major function of ABMs because it helps understand how simple rules generate complex structures. ABMs' explanatory power is highly generative, especially in social sciences due to the fact that it explains which macro-structures such as epidemic dynamics or social evolutions emerge in population of heterogeneous agents that interact locally and in non-trivial way under a set of tenable behavioral rules(Epstein, 2008).

ABMs facilitate the *experimentation process*(Leal & Napoletano, 2017). They can be run repeatedly to discern variations in their dynamics and in their outputs (Wilensky & Rand, 2015). some models show a very little variations during several runs. Some have a path-dependency nature(Brown, Page, Riolo, Zellner, & Rand, 2005) and some exhibit tremendous variations from run to run. Through experimentation, system modelers get informed of how input parameters affect model's outputs. Therefore, they can make various scenarios for achieving the targeted behavior.

ABMs are sometimes used for *prediction* purposes. Subject matter experts frequently use models to get a picture about possible future states. Like every model the quality of ABMs' prediction relies on the accuracy of its input parameters and since society is a complex system with an unspecified degree of uncertainty and very high sensitivity to small-scale events, no prediction can be deemed as absolutely right (Moss, 2008; Wilensky & Rand, 2015). prediction differs from description where the modeler describes the past or present states of the system, for example when a modeler describes what changes first occurred in the system. Moreover, prediction is also district from explanation, for example Plate tectonics definitely explains earthquakes, but does not help us to predict the time and place of their occurrence or evolution is commonly accepted as explaining speciation, but it is impossible to predict next year's flu strain(Epstein, 2008). Nonetheless, when subject matter experts claim to have used ABMs for purpose of prediction, they actually use ABMs either for description or explanation (Wilensky & Rand, 2015).

ABMs has a high functionality for *education and analysis* (Blikstein & Wilensky, 2009; Sengupta & Wilensky, 2009; Wilensky & Reisman, 2006). Educators can develop models for people that they have never seen before. For example, educators can model some examples of mutualism between individuals of different species when both individuals benefit[4]. Moreover, models can simulate a system that may not be readily available from real-world observations, therefor they can be very thought-provoking and enable learners to go beyond their observations and conduct experiments just like scientists.

when a subject matter expert is going to gain a deeper understanding of a phenomena about which there is not enough theory, *thought experiment* can be very useful. Though experiment is another suitable area for ABMs. This type of experiment is done to achieve its purpose without benefit of execution(Sorensen, 1998). Thought experiment is conducted when the real-world experiments are neither affordable nor possible to execute (Rangoni, 2014). It has a wide application in social and natural sciences. Through this method, researchers can get aware of the logical consequences of their hypotheses. For example, what will happen if a personnel of a company all telework in odd days of a weak? ABMs can be very useful in thought experiments especially when people want to deal with complex systems such organization and society. Such systems are far from a real-word laboratory where it is possible to control some variable (as control group) and measure the effect of test on other variables (as treatment group). As a matter of fact, in such systems, there are numerous causal factors that are mainly interdepended over which we have on or a very limited control (Savona, 2005). So real-world experiments can rarely be executed in such systems. This has led researchers of social fields to utilize the potential of though experiment in simulating the consequences of their hypothesized mechanism.

---

[4] An interesting example can be the mutualism between a goby and a shrimp. The shrimp digs a burrow in the sand and cleans it up where both species can live. Since the shrimp is almost blind, it has a high vulnerability to predators outside the burrow. When the shrimp is under dangerous conditions the goby goes over to warn the shrimp by touching it with its nail. This causes both the shrimp and goby quickly back into the burrow (Helfman, Collette, Facey, & Bowen, 2009).

## 6.3.    Modeling approaches to ABM

In ABM literature, modeling approaches can be divided based on a number of aspects. Three of the most important aspects include 1-goal of modeling, 2- development method 3- elaboration strategy. In terms of modeling goal, ABMs can be grouped into two major categories of *phenomena-based modelling and exploratory modelling* (Wilensky & Rand, 2015). In phenomena-based modeling researchers begin with a known target phenomenon. Typically, that phenomenon has a characteristic pattern, known as a reference pattern. Examples of reference patterns might include common housing segregation patterns in cities, diffusion of a specific ICT technology, spiral-shaped galaxies in space, leaf arrangement patterns on plants, or oscillating population levels in interacting species (Wilensky & Rand, 2015). These reference patterns are those statistical regularities that econometricians suppose as stylized facts for example, the way price affects supply or demand. The goal of phenomena-based modeling is to create a model that will somehow capture the reference pattern. In ABM, this translates to finding a set of agents, and rules for those agents, that will generate the known reference pattern. Once you have generated the reference pattern you have a candidate explanatory mechanism for that pattern and may also vary the model parameters to see if other patterns emerge, and perhaps try to find those patterns in data sets or by conducting experiments. Phenomena-based modeling can also be used with other forms of modeling, such as equation-based modeling. In equation-based modeling, this would mean writing equations that will give rise to the reference pattern. Evidently all empirical validations perform well in case of phenomena-based modelling where there is a reference pattern against which the accuracy of model's results is measured.  The second core modeling form is exploratory modeling. This form is perhaps less common in equational contexts than it is in ABM. In exploratory modeling with ABM, a researcher can create a set of agents, define their behavior, and explore the patterns that emerge. One might explore them solely as abstract forms, much like cellular automata developed by Conway (Conway, 1976) But to count as modeling, we must note similarities between the behavior of our model and some phenomena in the world just as patterns generated by cellular automata like oscillators and spaceships (Wolfram, 1983). Then ABM should be refined our model in the direction of perceived similarities with these phenomena and converge toward an explanatory model of some phenomenon. Phenomena-based modelling stems from the notion that there is an objective and real but unobservable data generating mechanism and that the purpose of any model is to represent elements of that mechanism in ways that generate some of the same data. But, in the case of exploratory based modeling the purpose of the models is the representation of perceptions by policy analysts and other stakeholders in the relevant social processes(Moss, 2008). The phenomenon based modeling is a class of agent based models that has much in common with mainstream economic models. They incorporate utility functions; they employ numerical representations of phenomena and attributes naturally described in qualitative terms by the individuals being represented and by other stakeholders. The exploratory modeling is a class of models emerging from a process that is embedded in the social process of policy and strategy formation. Such models are typically couched in linguistic terms (i.e. mental models) used by stakeholders rather than numerical variables convenient and meaningful only to modelers. The models are developed to facilitate stakeholder participation in the model design and validation process. They are intended precisely to represent the perceptions of stakeholders in order to bring clarity to scenarios built to explore the possibilities - the opportunities and dangers - of an uncertain future. The major function of exploratory modeling is to enable the subject matter experts (SMEs) to see what outcomes their mental model(s) can produce when implemented in a real-world system. Therefore, they can have lots of advantages for those who deal with understanding complex systems. In terms of elaboration strategy, ABMs can be grouped into KISS, KIDS and TAPAS. KISS strategy that stands for "Keep It Simple, Stupid". This notion is rooted in the Occam's razor principle stating "while being faced with a number of competing hypotheses for a problem, one should select a hypothesis that has a fewer assumption" This principle advocates the law of parsimony and agent-based modelers that use this principle "start from simple models and gradually sophisticate it to answer their question". KIDS strategy which stands for "Keep It Descriptive, Simple" is in opposite direction of KISS strategy. Advocates of KIDS strategy start from descriptive models and gradually simplify them to answer their questions.(Elsenbroich & Gilbert, 2014; Windrum et al., 2007)[5]. TAPAS strategy that stands for (Take A Previous model and Add Something) .In this strategy, modelers take an existing model and successively modify it through adding new features or relating initial assumption(Frenken, 2006).

In terms of development, ABMs can be grouped into two major categories of *theory-based modelling and evidence-based modelling* (Moss, 2008). ABMs can be developed via a theory. Actually a theory that specifies the behavioral rules of agents or the statistical regularities that the model is designed to explain them. Since theory-based ABMs are built upon prior studies (essentially empirical ones) and aimed at stimulating real-world aggregates such as technology diffusion, disease spread and inflation formation, they are frequently in phenomena-based modeling where the modelling of a real-world pattern is the purpose of the modeler. Besides theories, ABMS can also be developed based on evidence. The evidence-based modeling is used when researchers have a mental model concerning behavioral rules of agents of a system and they are interested in understand the collective behavior of that system when its agents interact with each other. As it can be inferred, this approach is the foundation of exploratory modeling approach where a model is developed for representing the emergent properties of researcher's mental models. Evidence-based ABMs can be developed either as participatory simulations or individual thought experiments. Participatory simulations approach is useful for modeling systems that there is not enough data about them. According to this approach, an agent-based model is directly developed through direct participation of stakeholders of the problem. This modeling is very useful in research and education(Colella, 2000; Frey & Goldstone, 2013). One of major forms of participatory simulation is "companion modeling" (ComMod) which is an iterative participatory approach where multidisciplinary researchers and stakeholders work together continuously throughout a three-stage cycle field work-> modelling -> simulation -> field work again (Barreteau, 2003). ComMod follows two basic objectives. First it increases the understanding of complex systems through its three-stage cycle. second it supports collective decision-making Processes in Complex Situations. In this case, the approach facilitates collective these kinds of processes by making more explicit the various points of view and subjective criteria, to which the different stakeholders refer implicitly or even unconsciously. Indeed, as demonstrated in past research(Funtowicz & Ravetz, 1994) when facing a complex situation, the decision-making process is evolving, iterative, and continuous. It means that this process produces always imperfect "decision acts", but following each iteration they are less imperfect and more shared. Principally, the main principle of the ComMod approach is to develop simulation models integrating various stakeholders' points of view and to use them within the context of platforms for collective learning. This is a modeling approach in which stakeholders participate fully in the construction of models to improve their relevance and increase their use for the collective assessment of scenarios. The general objective of ComMod is to facilitate dialogue, shared learning, and collective decision-making through interdisciplinary and "implicated" action-oriented research to strengthen the adaptive management capacity of local communities (Barreteau, 2003). As a software engineering method Virtual Overlay Multi Agent System (VOMAS) has been used to improve ComMod methodology (Niazi & Hussain, 2012).VOMAS is used for facilitating last two stages of ComMod, namely modeling and simulation. This method has been very successful in building verified and validated agent based models (Niazi, Muaz A; Hussain, Amir; Kolberg, 2017).Exploratory ABM has a wide potentiality for thought experiment (Rangoni, 2014). As discussed above, this approach is grounded on theory based modelling and is used when a SME or a team of SMES like to discern what will possibly emerge out of their mental models before being executed in real world. Outcomes produced by an exploratory ABM can find a number of empirical supports in real world data. Therefore, it can also play a vital role in theory development. (Macy & Willer, 2002). In this paper, an exploratory ABM has been used to show (1) how economic inequality emerges within an economic system (2) how Islamic charity (sadaqah) and allocation strategies of charity entities can help reduce this emergent economic inequality.

---

[5] Technically, KISS and KIDS can be supposed as two opposite ends of spectrum. In an effort for developing a unified strategy, Rand and Wilensky (2007) developed full spectrum modeling strategy. In this strategy models are developed in a progressive way ( either from simple to descriptive or s descriptive to simple) and the phenomenon under study is modelled at multiple levels of details(Rand & Wilensky, 2007).

## 7. Model specifications and simulation

For representing how sadaqah can help reduce the economic inequality of a society, two economic systems have been simulated using an exploratory agent-based modeling approach. The firs system (system I) is mainly inspired by the work of Wilensky and Rand (2015). This model has been programmed in Netlogo programming language. As one of the most frequently used agent based modeling languages, Netlogo was developed by Uri Wilensky in 1999. Science its developments, it has been regularly updated in sequence of versions and a number of extensions. Readers can refer to Netlogo home page ccl.northwestern.edu/netlogo/ in order to get more information of this agent based modeling language. However, to make this model more suitable for this study, the authors of this paper have added some new features to its original version including 1- the possibility of changing the number of agents in the interface view, 2- the possibility of choosing money of agents in the interface view, 3-The possibility of showing the amount of money of each agent by its color in the world view so that the richer agents have a darker color and go north wise while the poorer agents have a lighter color and go south wise (figures 1 and 2) and 4-The possibility of simulating when agents start with unequal amount of money in initial condition (figure 2). The second system (system II) is an extension of system I which includes a number of new features such as the ability of human agents to give sadaqah, the establishment of charity agents, the strategies that charity entities can take to allocate the sadaqah among the needy agents and so on. In order to make the replication of this model easier, the agent-based simulation of system II can be directly searched and downloaded from **http://modelingcommons.org/account/login** that is a useful platform for communicating and discussing agent-based models written in Netlogo[6]. The conceptual models (textual models) of system I and system II along with their simulation and analysis would be discussed in the following.

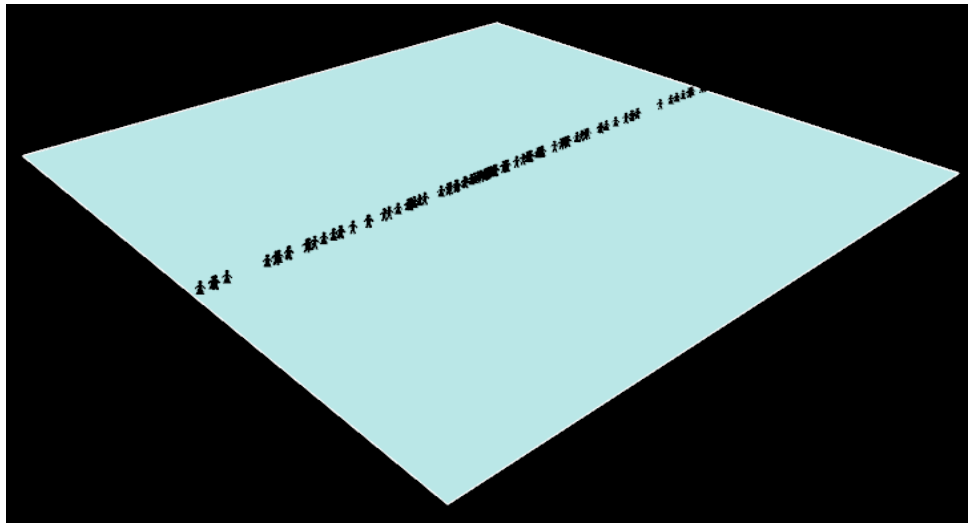

Figure 1: A 3D view of 100 Agents with equal amount of money in the initial conditions

---

[6] It should be searched with the title of "Economic inequality and Islamic charity".

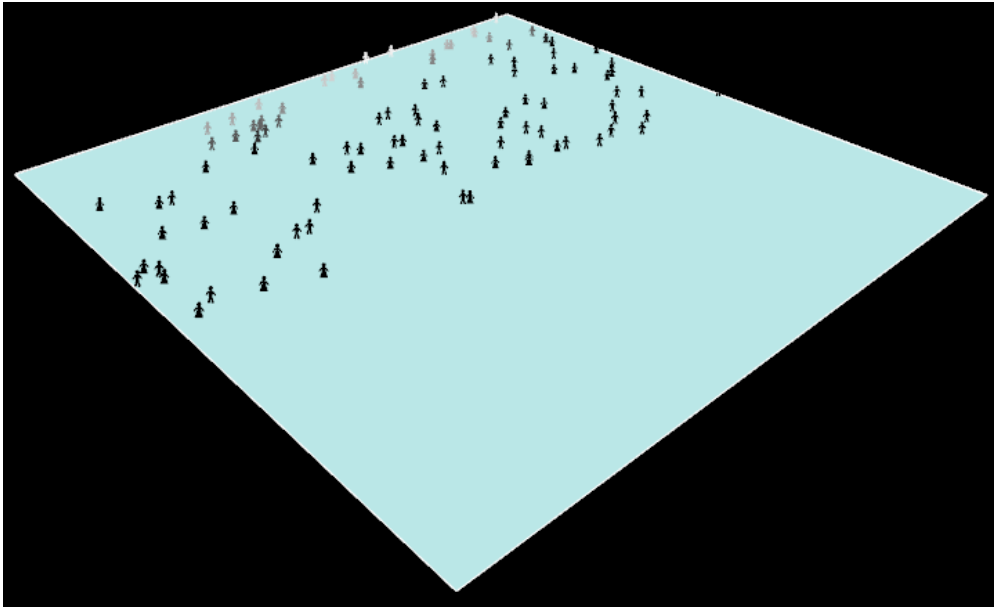

Figure 2: A 3D view of 100 Agents with unequal amount of money in the initial conditions

## 7.1. System I:

This system is built upon five assumptions including (1) there is a society in which N number of persons live, (2) there is a global clock that shows the time by each tick and it is completely discrete, (3) each person has M amount of money that can be equal or unequal in the initial conditions, (4) there is a money gap between the five low deciles (bottom 50%) and the tenth decile (top 10%). It can be positive, zero or negative. when it is positive, it shows the money of five low deciles is more than that of tenth decile. In case of being zero, it shows the money of five low deciles is equal to that of tenth decile and if it is negative, it indicates the money of five low deciles is less than that of tenth decile, (5) there is a critical threshold which shows the criticality of economic situation when the value of money gap becomes lower than it. According to this system, if all people (N = 500) of the society have an equal amount of money (M =100) in the initial conditions and donate a unit of money to each other randomly and by each time tick as long as each person's money is more than zero, what would be the answers to following questions:

1- How will the probability distribution of money be in tick of 100?
2- How will the probability distribution of money be in tick of 1000?
3- How will the probability distribution of money be in tick of 9000?

Because of the stochastic nature of agent-based models, the probability distribution of money in system I has been simulated in 10 different runs[7]. In run 1, According to figure 3, After 100 ticks, the probability distribution of money is similar to a normal distribution as M ~ N (100, 110.612). This distribution shows that money of each person is very inclined to the average money of society. Therefore, all deciles of society have a somewhat similar amount of money. As shown in figure 6, there is a great distance between the money volume of all five lower deciles (bottom 50% with 22885 units of money) and that of tenth decile (top 10% with 5933 units of money) meaning that bottom 50% has 16952 units of money more than top 10%.

---

[7] - All runs are experimented in seeds of 1000, 2000,3000,4000,5000,6000,7000,8000,9000 and 1000 respectively.

As demonstrated in figure 4, the probability distribution of money has become flatter when the system I is in tick of 1000. This normal distribution has a mean of 100 and variance of 972.661 implying that the majority of the society have a money inclined to 100 units of money and just a few of them own a very high amount of money. However, regarding figure 7 it can be concluded, the total amount of money of bottom 50% has become mitigated to 18835 while the top 10% has accumulated a remarkable amount of money 7783 and money distance has decreased to large degree 11052.

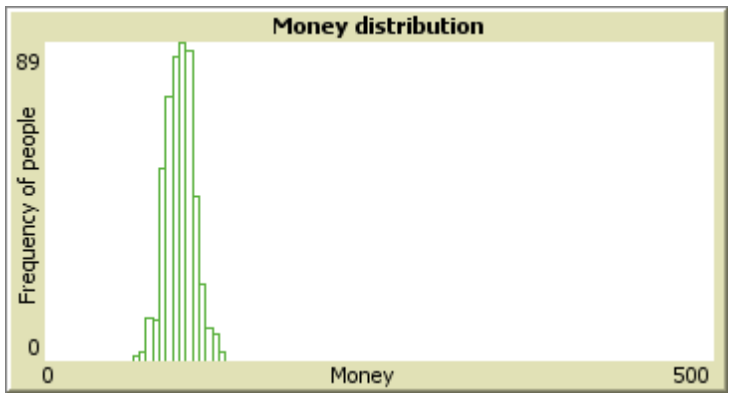

Figure 3: probability distribution of money in tick 100

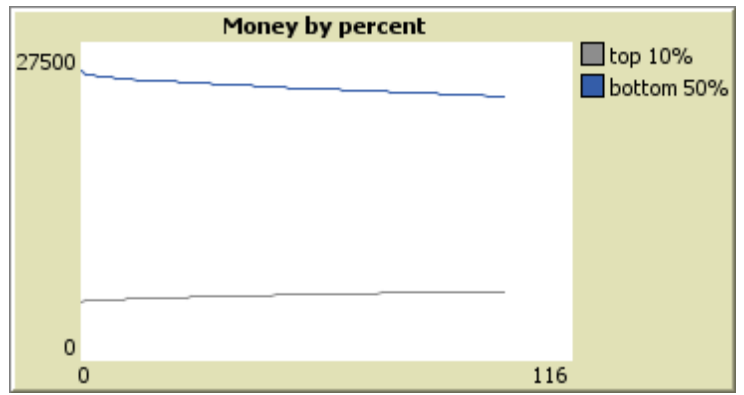

Figure 6: distance of bottom 50% from top 10% in tick 100

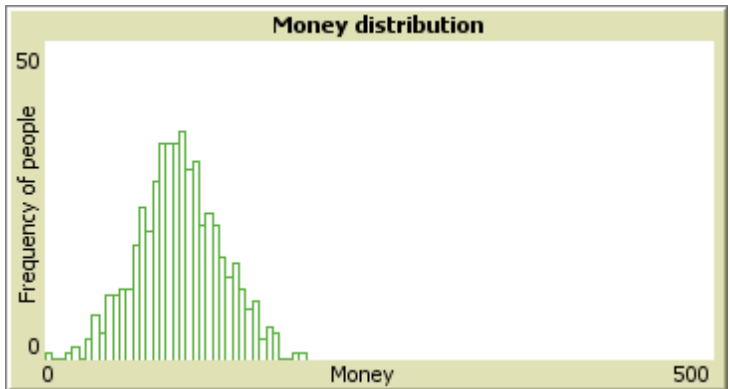

Figure 4: probability distribution of money in tick 1000

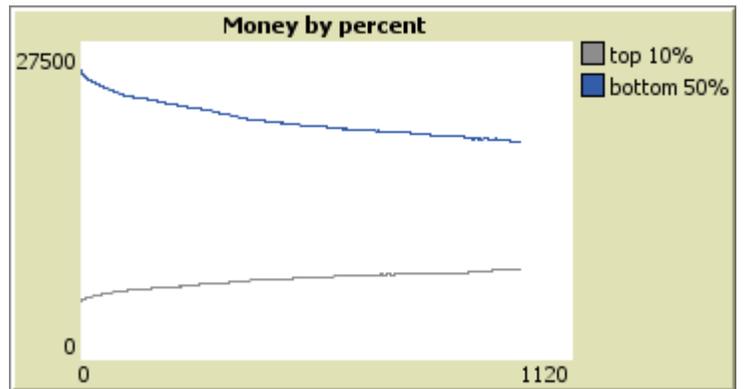

Figure 7: distance of bottom 50% from top 10% in tick 1000

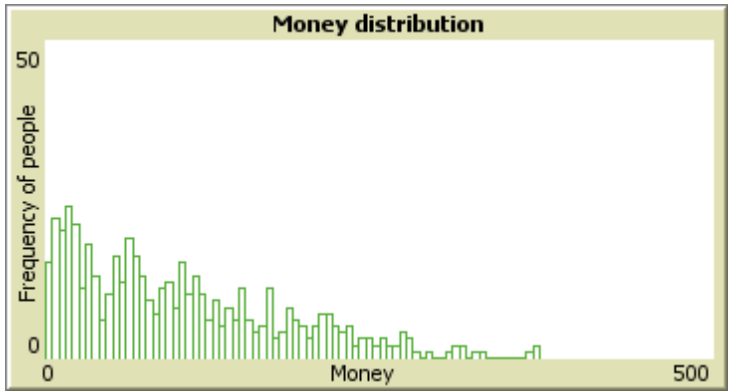

Figure 5: probability distribution of money in tick 9000

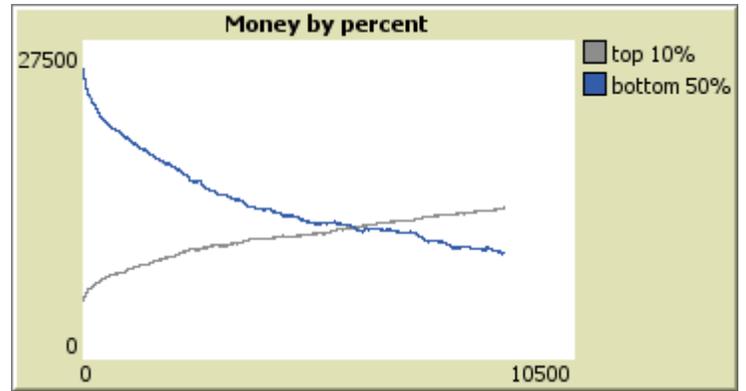

Figure 8: distance of bottom 50% from top 10% in tick 9000

According to figure 5, when system I comes to tick of 9000, it shows a Boltzman-Gibbs distribution implying that just a very few of persons have accumulated a great amount of money and a large number of them have just gained a little amount of money. In this case the bottom 50% has an amount of money of 9134 while top 10% has an amount equal to 13052 that is 3918 units more than that of bottom 50%. The money volume of top 10% has exceeded that of bottom 50% in tick of 5800 since then the gap has increased (figure 8). Statistical summary of all 10 runs for ticks of 100, 1000 and 9000 is described in tables 1, 2 and 3 respectively.

Table 1: system up to tick of 100

| # of runs | μ | Variance | Total money of top 10% | Total money of bottom 50% | diff | Tick of critical stage |
|---|---|---|---|---|---|---|
| Run 1 | 100 | 110.621 | 5933 | 22885 | 16952 | - |
| Run 2 | 100 | 104.561 | 5877 | 22959 | 17082 | - |
| Run 3 | 100 | 102.997 | 5908 | 22988 | 17080 | - |
| Run 4 | 100 | 92.641 | 5824 | 23114 | 17290 | - |
| Run 5 | 100 | 100.917 | 5867 | 22974 | 17107 | - |
| Run 6 | 100 | 97.002 | 5850 | 23025 | 17175 | - |
| Run 7 | 100 | 89.254 | 5840 | 23132 | 17292 | - |
| Run 8 | 100 | 90.509 | 5884 | 23149 | 17265 | - |
| Run 9 | 100 | 100.044 | 5913 | 23028 | 17115 | - |
| Run 10 | 100 | 106.052 | 5963 | 22987 | 17024 | - |

Table 2: system up to tick of 1000

| # of runs | μ | variance | Total money of top 10% | Total money of bottom 50% | diff | Tick of critical stage |
|---|---|---|---|---|---|---|
| Run 1 | 100 | 972.661 | 7783 | 18835 | 11052 | - |
| Run 2 | 100 | 950.793 | 7578 | 18871 | 11293 | - |
| Run 3 | 100 | 1031.490 | 7926 | 18555 | 10629 | - |
| Run 4 | 100 | 1092.669 | 7763 | 18279 | 10516 | - |
| Run 5 | 100 | 1054.322 | 7837 | 18425 | 10588 | - |

| Run 6 | 100 | 986.492 | 7645 | 18778 | 11133 | - |
| Run 7 | 100 | 1001.531 | 7975 | 18765 | 10790 | - |
| Run 8 | 100 | 1259.535 | 8191 | 17949 | 9758 | - |
| Run 9 | 100 | 958.436 | 7712 | 18822 | 11110 | - |
| Run 10 | 100 | 1063.314 | 7840 | 18480 | 10640 | - |

Table 3: system up to tick of 9000

| # of runs | μ | variance | Total money of top 10% | Total money of bottom 50% | diff | Tick of critical stage |
|---|---|---|---|---|---|---|
| Run 1 | 100 | 6256.793 | 13052 | 9134 | -3918 | 5800 |
| Run 2 | 100 | 5359.6312 | 12692 | 10807 | -1885 | 6476 |
| Run 3 | 100 | 5042.136 | 12207 | 10878 | -1329 | 6310 |
| Run 4 | 100 | 4959.899 | 11890 | 10662 | -1228 | 6919 |
| Run 5 | 100 | 5443.066 | 12562 | 10225 | -2337 | 5801 |
| Run 6 | 100 | 4869.382 | 11871 | 10737 | -1134 | 6941 |
| Run 7 | 100 | 5597.174 | 12459 | 9723 | -2736 | 5520 |
| Run 8 | 100 | 5326.837 | 12560 | 10373 | -2187 | 5752 |
| Run 9 | 100 | 5533.523 | 12648 | 10046 | -2602 | 2725 |
| Run 10 | 100 | 5469.022 | 12428 | 10011 | -2417 | 5668 |

As a fundamental law of equilibrium statistical mechanics, Boltzman-Gibbs law states that any conserved quantity in a big system should follow an exponential probability distribution. According to Boltzman-Gibbs law, in a closed economic system, the total amount of money is conserved because it is not manufactured, consumed or destroyed. Thus the equilibrium probability distribution of money p(m) should follow the Boltzman-Gibbs law as equation 4:

$$P(m) = Ce^{-m/T} \tag{4}$$

Here m is money, C is a normalizing constant and T is an effective temperature equal to the average amount of money per agent. Some studies have shown that the money distribution follows a Boltzman-Gibbs law when it is conserved and exchanged in a closed system. According to this property of money distribution a very few of persons will accumulate a great amount of money while a large number of them just gain a little amount of money (Dragulescu & Yakovenko, 2000; Ferrero, 2004; Yakovenko & Rosser, 2009). As it can be seen in table 3, in all ten runs, the top 10% has outweighed the bottom 50% in terms of accumulated money and consequently economic situation entered a critical stage in all runs. Thus, in order to prevent this economic system from becoming more critical, system I has been extended into system II by adding some more mechanisms to it.

## 7.2. System II:

M number of charity organizations are added to system I. These organizations come to scene when the economic situation becomes critical (i.e., the money of 10% of the society becomes equal to or more than that of its 50%). The mission of these organizations is to help the low five deciles not to deteriorate more in the depth of poverty. These beneficiaries work based on the sadaqah that benefactors give to them in order to distribute it among five lower deciles. in terms of sadaqah-giving and distribution, these charities can take one of three allocation strategies of A, B or C. Strategy A is applied when just the richest person of the society gives a unit of his or her money to the charity organization and it allocates that money to the poorest person of the society. When c% of members of decile 10 give a unit of money to the charity organization and it allocates that of amount of money among d% of members of five lower deciles, the charity organization has used strategy B. The charity entities use strategy C when k% of members of decile 10, p% of members of decile 9 and v% of members of decile 8 give money (everybody one unit of money) to them and they distribute that amount of money among x% of members of decile 1, y% of members of decile 2 and z% of members of decile 3 respectively. This model can help answering the following questions:

1- How will strategy A affect the economic system when it enters a critical stage?
2- How will strategy B affect the economic system when it enters a critical stage?
3- How will strategy C affect the economic system when it enters a critical stage?

According to table 3, for run 1, the system has entered the critical stage in tick of 5800. The charity organization has used strategy A to help system exit this stage but it has not got out of the critical stage in the next ticks (up to 9000). It means that as long as the charity entity uses strategy A to help system get out of the critical stages, it fails to exit and returns to critical stage by each tick. The return period or recurrence interval of the critical stage is a key indicator for measuring the sustainability of allocation strategies. The visualization of how charity organization uses strategy A in tick of 5800 is shown in figure 9.

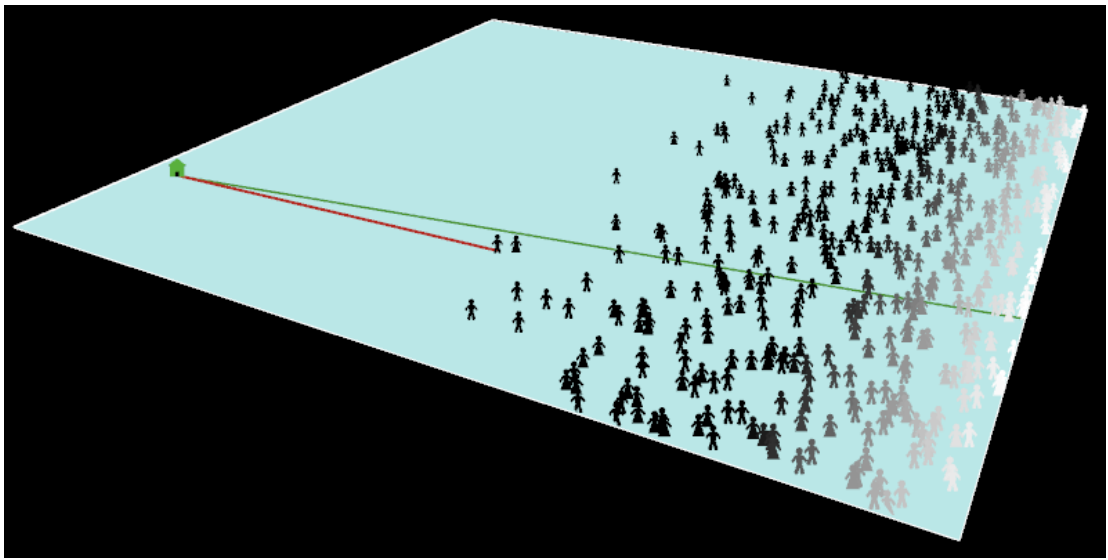

Figure 9 –A 3D view of distribution of sadaqah by charity entity while using strategy A in tick 5800

According to figure 9, the red line indicates sadaqah from the richest person of the society (benefactor) to charity and the green line indicates the sadaqah distributed by charity to the poorest person of the society. As it can be noticed, the spatial position and color of rich persons are different from those of the poorer ones. Such differences are because of a mechanism embedded into this model which forces the rich get darker color and move north wise while making the poor get brighter and move south wise. As shown in figures 10 to 15, for run 1, strategy A, strategy B (with parameters of c=100 and d= 20) and Strategy C (with parameters of k = 100, p = 60, v= 40, x= 100, y=60 and z= 40) have been applied for handling the economic critical stages up to 9000 ticks.

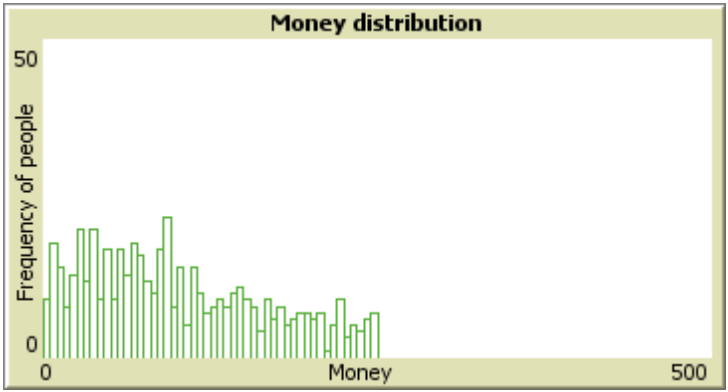

Figure 10 -Distribution of sadaqah by charity entity while using strategy A for 9000 ticks

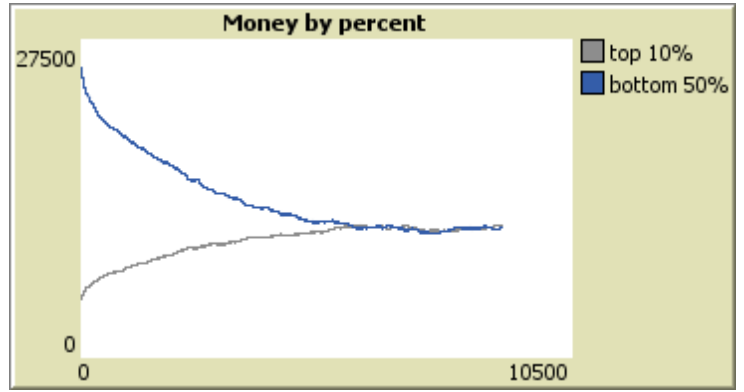

Figure 13: distance of bottom 50% from top 10% in tick 9000 when the charity entity uses strategy A

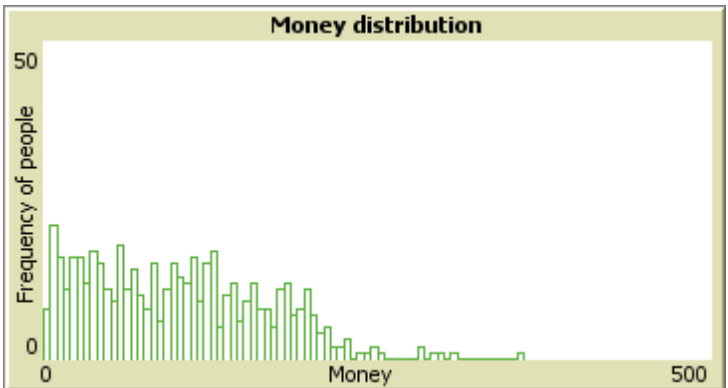

Figure 11 -Distribution of sadaqah by charity entity while using strategy B for 9000 ticks

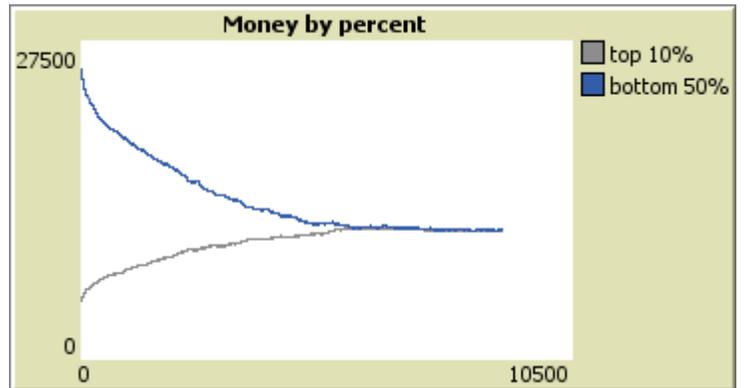

Figure 14: distance of bottom 50% from top 10% in tick 9000 when the charity entity uses strategy B

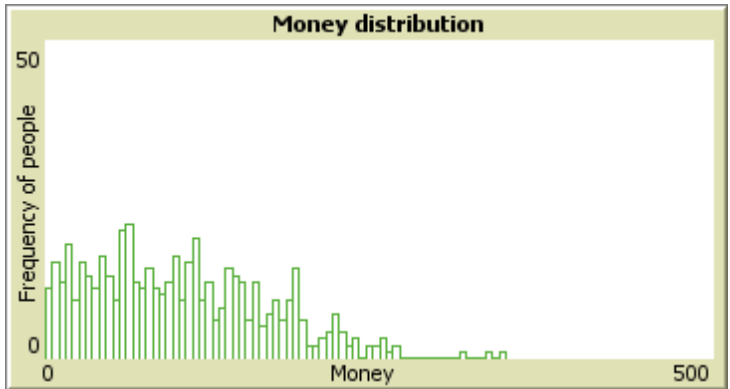
Figure 12 -Distribution of sadaqah by charity entity while using strategy C for 9000 ticks

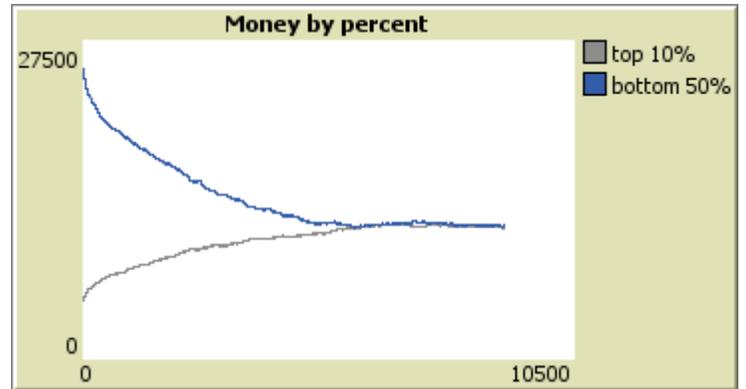
Figure 15: distance of bottom 50% from top 10% in tick 9000 when the charity entity uses strategy C

As demonstrated in figures 10, 11 and 12, by each of allocation strategies, charity organization have tried to help the economic system exit the critical stage while entering it in every time. In this case when the total money of top 10% becomes equal to or exceeds that of bottom 50%, the total money of top 10% will be forced back to a value less than that of bottom 50% depending on how many of three higher deciles participate in giving sadaqah and also which allocation strategies the charity organizations use to distribute the Sadaqah among poorer deciles. According to the Boltzman-Gibbs law after one of allocation strategies is implemented the money (as a conserved quantity here) will tend to become accumulated into hands of a very few of people. Thus the systems enter the critical stage again. The number of times that economic system returns to the critical stage after a typical strategy is used for resource redistribution in it, is a key factor for measuring how sustainable that strategy is.

As it can be seen from figures 13,14, 15, when a typical strategy is used, the society will have a specific form of money distribution in the final tick (here 9000). Therefore, the less variance the distribution has, the better resource allocation strategy has been used. Because it has more reduced the economic inequality among people (in terms of money distribution). Thus, the variance of money distribution in the society when it enters the final tick is another key factor for measuring the efficiency of the resource allocation strategy. The number of return periods of critical stages and variance of the money distribution in all runs have been presented for all of strategies in tables 4, 5 and 6.

Table 4: application of strategy A up to 9000 ticks

| # of runs | # of return periods | Money of top 10% | Money of bottom 50 % | Variance | diff |
|-----------|---------------------|------------------|----------------------|----------|------|
| Run 1 | 2015 | 11328 | 11219 | 4465.974 | -109 |
| Run 2 | 1517 | 10792 | 10888 | 4398.300 | 96 |
| Run 3 | 2287 | 11121 | 10784 | 4596.4128 | -337 |
| Run 4 | 1386 | 10815 | 10650 | 4516.769 | -165 |
| Run 5 | 2150 | 11001 | 10838 | 4489.615 | -163 |
| Run 6 | 1656 | 11208 | 11314 | 4341.547 | 106 |
| Run 7 | 2638 | 10745 | 10716 | 4479.010 | -29 |
| Run 8 | 1590 | 11383 | 11298 | 4419.070 | -85 |
| Run 9 | 2368 | 10725 | 10364 | 4624.541 | -361 |
| Run 10 | 2059 | 10997 | 10727 | 4459.478 | -270 |

Table 5: application of strategy B up to 9000 ticks

| # of runs | # of return periods | Money of top 10% | Money of bottom 50 % | Variance | diff |
|---|---|---|---|---|---|
| Run 1 | 42 | 11032 | 11182 | 4399.110 | 150 |
| Run 2 | 47 | 11353 | 11459 | 4415.555 | 106 |
| Run 3 | 48 | 11475 | 11592 | 4443.134 | 117 |
| Run 4 | 18 | 11357 | 11474 | 4364.448 | 117 |
| Run 5 | 34 | 11256 | 11316 | 4367.595 | 60 |
| Run 6 | 26 | 11594 | 11673 | 4378.737 | 79 |
| Run 7 | 37 | 11243 | 11351 | 4377.206 | 108 |
| Run 8 | 37 | 11213 | 11272 | 4426.685 | 59 |
| Run 9 | 39 | 11351 | 11420 | 4435.002 | 69 |
| Run 10 | 22 | 11252 | 11292 | 4379.559 | 40 |

Table 6: application of strategy C up to 9000 ticks

| # of runs | # of return periods | Money of top 10% | Money of bottom 50 % | Variance | diff |
|---|---|---|---|---|---|
| Run 1 | 31 | 11224 | 11520.3 | 4302.669 | 296.33 |
| Run 2 | 19 | 11681 | 11760 | 4388.751 | 79 |
| Run 3 | 28 | 11464 | 11544.3 | 4409.166 | 80.33 |
| Run 4 | 14 | 11305 | 11565.3 | 4252.689 | 260.33 |
| Run 5 | 30 | 11371 | 11519.7 | 4315.761 | 148.66 |
| Run 6 | 22 | 11667 | 11850 | 4394.670 | 182.99 |
| Run 7 | 24 | 11525 | 12082.7 | 4166.427 | 562.666 |
| Run 8 | 27 | 11579 | 11699 | 4363.750 | 119.99 |
| Run 9 | 26 | 11286 | 11379.7 | 4326.632 | 93.66 |
| Run 10 | 27 | 11635 | 11707.7 | 4311.794 | 72.66 |

According to tables 4,5 and 6, when strategy A is used, the average number of return periods is 1966.6 (roughly 1967) times Meaning that in all 10 runs, the economic system has returned to critical stage with the average of 1967 times. By applying strategy B, the system has shown a remarkably small average number of return periods which is 35. This number shows the second strategy is far more sustainable than strategy A in helping the system exit the critical stage for long time intervals. Strategy C has shown the number of 24.8 (roughly 25) for average of return periods in all runs.

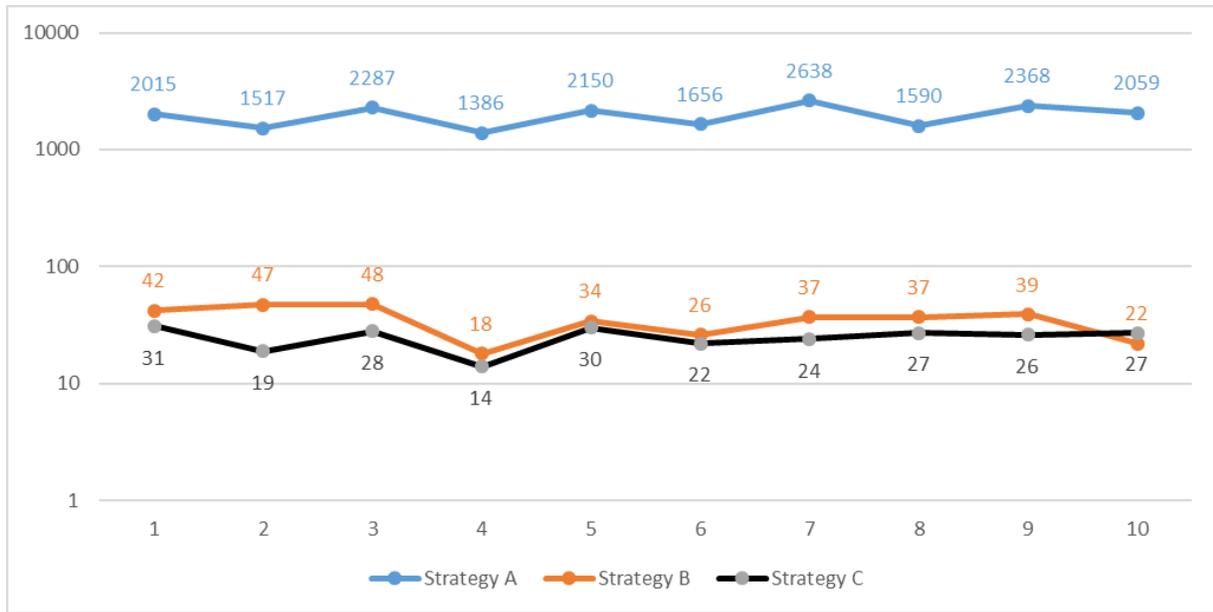

Figure 16: The sustainability of strategies

The number of return periods for each strategy has been visualized in figure 16. The vertical axis of this figure is the number of return periods that is made logarithmic (for better representation) and the horizontal axis stands for number of runs of simulation. As it can be seen, the strategy A has had the lowest level of sustainability while the strategy C has the highest one. Except run 10, strategy C has shown the lowest number of return periods for all other runs. Though this number doesn't significantly differ from that of strategy B, it shows more sustainability.

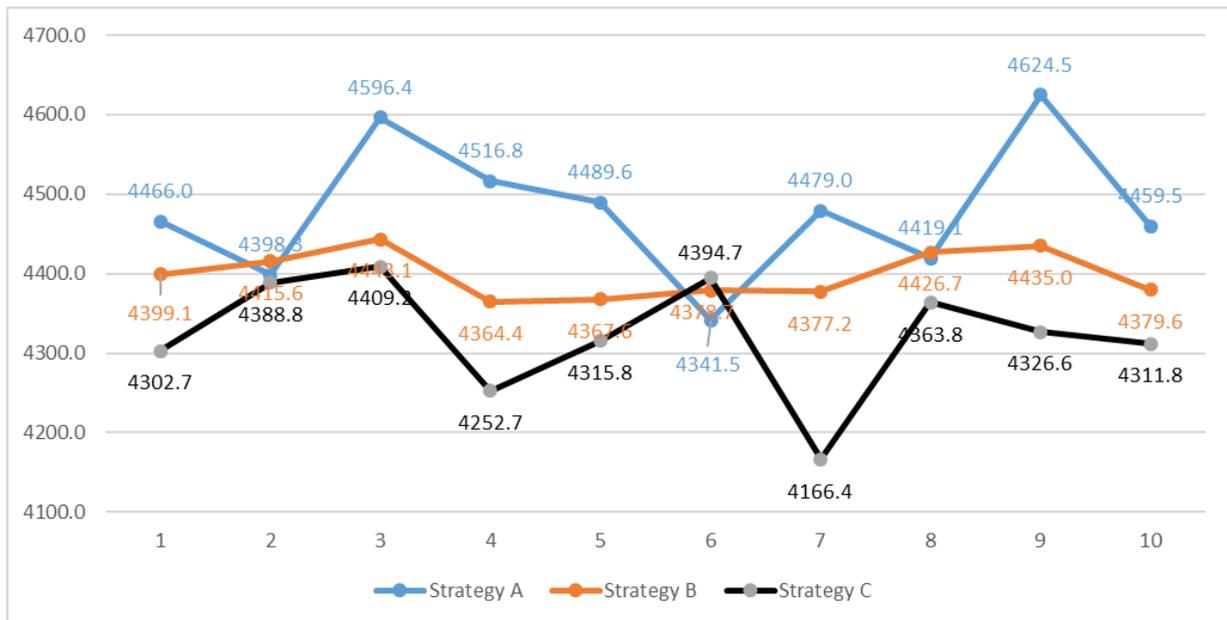

Figure 17: Variance values of strategies

Another information that has be inferred from tables 4 to 6 refers to how efficient each of allocation strategies has been in reducing the overall variance among people in terms of money distribution. This indicator shows how much agents differ from each other in terms of money volume. when strategy A is applied, the average overall variance of money distribution is 4624.5 meaning that in all 10 runs of simulation, the average overall variance of distribution of money in economic system has been 4479.07. by applying the strategy B, the average overall variance has decreased to 4398.70. While the sadaqah paid in strategy B is 50 times more than that of strategy A (when systems enter critical stage), the average overall variance has not decreased remarkably (only 80.37 units). Strategy C has shown a better performance in reducing the variance of money distribution almost in all runs (except run 6). This strategy has had the overall average variance of 4323.2 for all 10 runs showing the fact that it is the most efficient strategy. The variance values of money distribution for each strategy have been visualized in figure 17. The vertical axis of this figure is values of variance (for better representation) and the horizontal axis stands for number of runs of simulation. As it can be seen, the strategy C has had the lowest values of variances in all runs except run 6. Thus, this strategy is regarded to have the highest efficiency.

## 8. Conclusion

This paper has served two purposes. The first purpose was the explanation of how economic inequality emerges in an economic system. Results indicates that in a closed economic system when agents exchange the money, it tends to be accumulated in the hands of a very few number of agents over time. Therefore, the distribution of money in the society will follow a power law distribution. these results find empirical support from the field of equilibrium statistical mechanics (Dragulescu & Yakovenko, 2000; Yakovenko & Rosser, 2009). As a fundamental law of this field, Boltzman-Gibbs law states that in a closed economic system, the total amount of money is conserved because it is not manufactured, consumed or destroyed so any conserved quantity in a big system should follow an exponential probability distribution. The second purpose was to simulate how Islamic charity (sadaqah) and allocation strategies of charity entities can help reduce the economic inequality emerged in a system. The results showed that sadaqah is highly effective in reducing the gap among economic deciles. So the more people pay sadaqah the more they can decrease the gap. A part from sadaqah, the results imply that the way charity entities allocate the money among lower economic deciles (i.e., bottom 50%) is of paramount importance. In strategy A when system enters a critical stage, just one unit of money (as sadaqah) is transited from the richest agent to poorest one. In strategy B, that amount of money becomes fifty times lager in each transition but the overall average variance (in comparison to average number of return periods) has not had a remarkable decrease. The money is made 2 times large in strategy C but it shows a great decrease in overall average variance in contrast to that of strategy B. The main reason for this is not only because of money volume increase but also largely because of the ways money resources are taken from higher deciles and allocated to lower ones. However, like all scientific works, this work has a number of limitations. The first limitation is the simplicity of the economic system. So extending the model to more elaborated levels can be a good subject for future studies. For example, human agents can be made more intelligent. They can have memory capacity, abilities for production, consumption and destruction, educational level, tendency for entrepreneurship and so many other psychophysiological features. New agent breeds can also be added to the model. For example, banks, factories, venture capital funds and so on. Such an extension can yield very interesting results. For instance, when factories hire more of their staff among those educated agents that belong to lower economic deciles, venture capital funds define some priorities for poor agents possessing high entrepreneurial tendency and a professional interaction is set among agents, some behavioral patterns will surprisingly emerge both at micro-scale (agent-level) and macro-scale (system-level). The second limitation is that just distribution of money has been the subject of this study and distribution of wealth or income has not been simulated. So, extending the system in order to show distribution of wealth and income will be a great contribution that can be made via future studies.